\documentclass[twocolumn,epjc3]{svjour3}

\RequirePackage[T1]{fontenc}

\smartqed

\RequirePackage{graphicx}

\RequirePackage{mathptmx}
\RequirePackage{flushend}
\RequirePackage[numbers,sort&compress]{natbib}
\RequirePackage[colorlinks,citecolor=blue,urlcolor=blue,linkcolor=blue]{hyperref}

\usepackage{amsmath,amssymb} % for \text command

\usepackage[labelfont=bf]{caption} % for nice captions
\usepackage{color} % for colored text background
\usepackage{booktabs} % for tables: toprule, midrule, bottomrule
\usepackage{hyperref} % for urls and linked references
\usepackage{braket}
\usepackage{siunitx}
\usepackage{upgreek}
\usepackage{bm}

\usepackage{natbib,hyperref}
\usepackage{doi}
\usepackage{subcaption}
\usepackage{gensymb}

\newcommand{\kr}{^{83\text{m}}\text{Kr}}
\newcommand{\tr}{\text{T}_2}
\newcommand{\korrM}{\mathbf{P}}

\usepackage{ifthen}

\makeatletter
  {\par\addvspace{17pt}\small\rmfamily
   \trivlist
   \item[\hskip\labelsep\bfseries #1]}
  {\endtrivlist\addvspace{6pt}}
\makeatother

\journalname{Eur. Phys. J. C}

\begin{document}

\title{Using krypton-83m to determine the neutrino-mass bias caused by a non-constant electric potential of the KATRIN source}

\author{Moritz Machatschek\thanksref{e1,iap}}
\thankstext{e1}{e-mail: moritz.machatschek@web.de}

\institute{This work was largely conducted while the author was affiliated with Institute for Astroparticle Physics~(IAP), Karlsruhe Institute of Technology~(KIT), Hermann-von-Helmholtz-Platz 1, 76344 Eggenstein-Leopoldshafen, Germany\label{iap}}

\date{Received: date / Accepted: date}

\maketitle

\begin{abstract}
Precision spectroscopy of the electron spectrum of the tritium $\upbeta$\,decay near the kinematic endpoint is a direct method to determine the effective electron antineutrino mass. The KArlsruhe TRItium Neutrino (KATRIN) experiment aims to determine this quantity with a sensitivity of better than $\SI{0.3}{\electronvolt}$ ($90\,\%$~C.L.). An inhomogeneous electric potential in the tritium source of KATRIN leads to a distortion of the $\upbeta$\,spectrum, which directly impacts the neu\-tri\-no-mass observable. This effect can be quantified through precision spectroscopy of the conversion electrons of co-cir\-cu\-la\-ted metastable $\kr$. This work perturbatively describes the effect of the source potential on the recorded spectra, and thus establishes the leading-order observables of this effect.
\end{abstract}

\section{Introduction}
\label{intro}

While neutrino-oscillation experiments show that neutrinos have mass~\cite{fukuda1999measurement,ahmad2002direct}, the absolute mass scale of neutrinos remains unknown. The KArlsruhe TRitium Neutrino (KATRIN) experiment is the latest experiment in the field of direct neu\-tri\-no-mass measurements with the goal to determine the absolute neutrino-mass scale, currently providing the best laboratory neutrino-mass upper limit of 0.45\,eV/c$^2$ (90\,\% confidence level (C.\,L.))~\cite{KATRINKNM1-5}. KATRIN uses high-precision electron spectroscopy to perform a shape analysis of the tritium $\upbeta$\,spectrum near the kinematic endpoint at 18.6\,keV, where the imprint of a non-vanishing neutrino mass is the most pronounced. Any other influence on the $\upbeta$\,spectrum shape in this region has to be well understood, since it would otherwise lead to a systematic bias of the neutrino-mass measurement. Thus, reaching the target sensitivity of better than 0.3\,eV/c$^2$ at 90\,\% C.\,L. requires the detailed study of systematic measurement uncertainties.

One major uncertainty is linked to the electric potential inside the tritium source. Inhomogeneities of the potential lead to a distribution of starting energies of the $\upbeta$\,electrons and thus to a distortion of the $\upbeta$\,spectrum, which needs to be characterized in order to reduce the systematic bias in the neutrino-mass measurement.

To this end, conversion electrons from $\kr$ are used as nuclear standard. In dedicated calibration measurements, traces of gaseous $\kr$ are circulated alongside tritium in the 10\,m long source, such that inhomogeneities of the potential are observable as distortion of the selected mono-energetic $\kr$ lines.

The focus of this work is to derive the leading-order effects caused by the electric potential; in section~\ref{ch:KATRIN}, the statistical moments of the potential, which are direct observables in the $\kr$ measurement, and relations between these moments, given the special properties of the KATRIN source. In section~\ref{ch:TritiumSystematics}, the systematic biases of the observable neutrino mass and $\upbeta$\,spectrum endpoint as function of these moments. Section~\ref{ch:Supplements} contains relevant supplements, which go beyond the main discussion.

\section{The starting potential in the KATRIN model}
\label{ch:KATRIN}

In the following, the relevant components of the KATRIN model are described and the essential mechanism for incorporating the potential is identified. Then, the leading-order observables of the potential are derived and their connection to the potential shape is established. Natural units with $c=e=1$ are used througout the paper.

\subsection{The KATRIN model}
\begin{figure*}
    \centering
    \includegraphics[width=\textwidth]{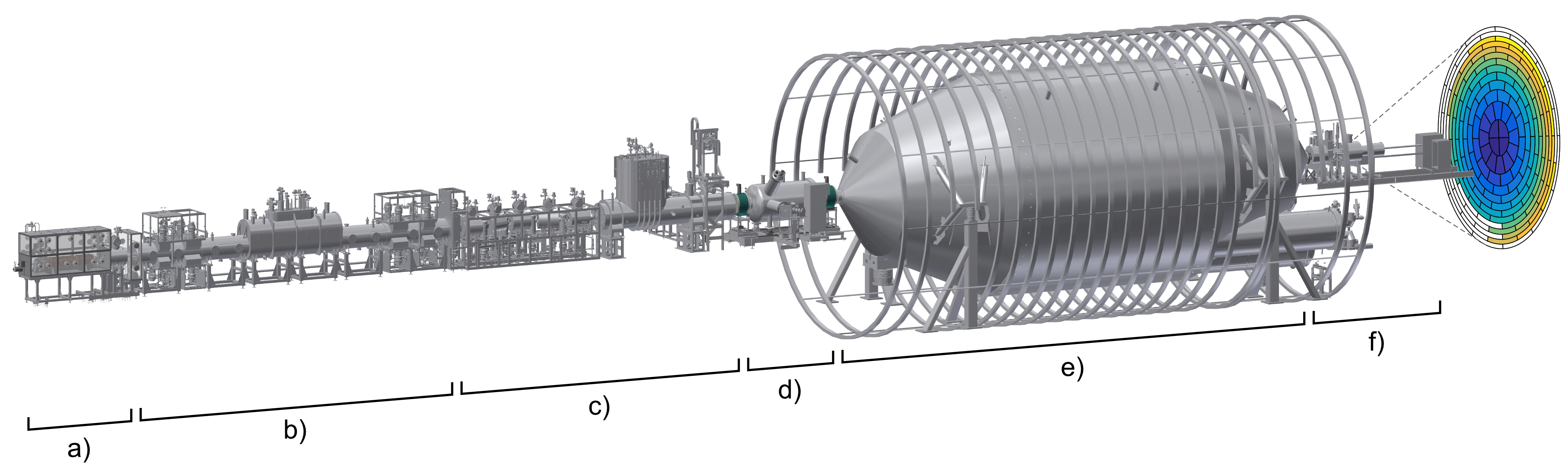}
    \caption{\textbf{Schematic of the KATRIN beamline with FPD.}
    The rear section a) is used for calibration and monitoring. The decay of gaseous tritium or krypton-83m takes place in the source b). Since the source is windowless, the gas is removed in the transport section c) using differential pumping and cryosorption, while the decay electrons are guided adiabatically to the spectrometer section d) and e) by a strong magnetic field. Electrons that can overcome the retarding potential of the main spectrometer e) are counted at the focal plane detector f). The segmentation of the focal plane detector in 148 pixels (shown on the right) and the magnetic guidance allow for radial resolution of the measurement, in particular of the source.}
    \label{fig:Beamline}
\end{figure*}
A schematic of the KATRIN experiment is shown in figure~\ref{fig:Beamline}. The reader is referred to~\cite{aker2021design} for a detailed description of the components and to~\cite{kleesiek2019upbeta} for details on the modeling.

Electrons are created by the $\upbeta$ decay of $\tr$ or internal conversion of $\kr$ inside the Windowless Gaseous Tritium Source (WGTS, figure~\ref{fig:Beamline}b)), which is a cylinder of 4.5\,cm radius and $L=10$\,m length. The electrons are guided through the experiment by a strong magnetic field. Radial variances in the physical quantities, like the electric potential of the axially-symmetric experiment, are mapped to distinct pixels of the Focal Plane Detector (FPD, figure~\ref{fig:Beamline}f)) by the magnetic field and are not considered in the following. However, longitudinal inhomogeneities (coordinate $z$) along one field line are integrated into a distorted spectrum, measured by one pixel. Determining this distortion caused by an unknown electric potential $V(z)$ in leading order is the goal of this paper. For ease of notation the $z$-integration is written as
\begin{equation}
    \int \frac{\mathrm{d}z}{L} \dots\equiv\Braket{\dots}~.
\end{equation}
in this section. Further integral operators are defined below using square brackets. Accordingly, since arguments in $\braket{\dots}$ and $[\dots]$ are always $z$-dependent functions, the $z$-dependency is often omitted. Arguments in $(\dots)$ are always scalars.

KATRIN uses an electrostatic spectrometer of the MAC-E filter (magnetic adiabatic collimation with electro-static filter) type (figure~\ref{fig:Beamline}e)) for the energy spectroscopy: only electrons (charge $q=-1$) with energies in forward direction larger than the retarding energy $qU$ (with retarding potential $U$) applied at the spectrometer can pass to the detector and get counted, all other electrons are reflected. Thus, the transmission condition $\mathcal{T}(E,\theta,U)$ is a function of the electron energy $E$, the retarding potential and the angle $\theta$ of the electron with regard to the magnetic field. By lowering the magnetic field inside the MAC-E filter by several orders of magnitude compared to the magnetic field in the source, the electron angles are collimated in forward direction. As a consequence, the transmission condition implicitly also depends on the ratio of the magnetic fields, as well as on higher order corrections like synchrotron radiation and relativistic effects among others. In order to avoid systematic effects, electrons with angles larger than $\theta_\mathrm{max}\approx51\degree$ are cut off. This is achieved using the magnetic mirror effect, by applying the strongest magnetic field directly in front of the detector.

The spectrum of the electrons gets modified by inelastic scattering of the electrons on the gas. Thereby the electrons lose the energy $\epsilon$ with a probability density $f(\epsilon)$. Multiple scattering events with multiplicity $i$ are described by convolving the so-called energy-loss function with itself
\begin{equation}
    f_i=\delta\underbrace{\otimes f \otimes f ...}_{i-\mathrm{times}}~,
    \label{eq:elossitimes}
\end{equation}
where $\delta$ is the $\delta$ distribution.
\begin{figure}
    \centering
    \includegraphics[width=.49\textwidth]{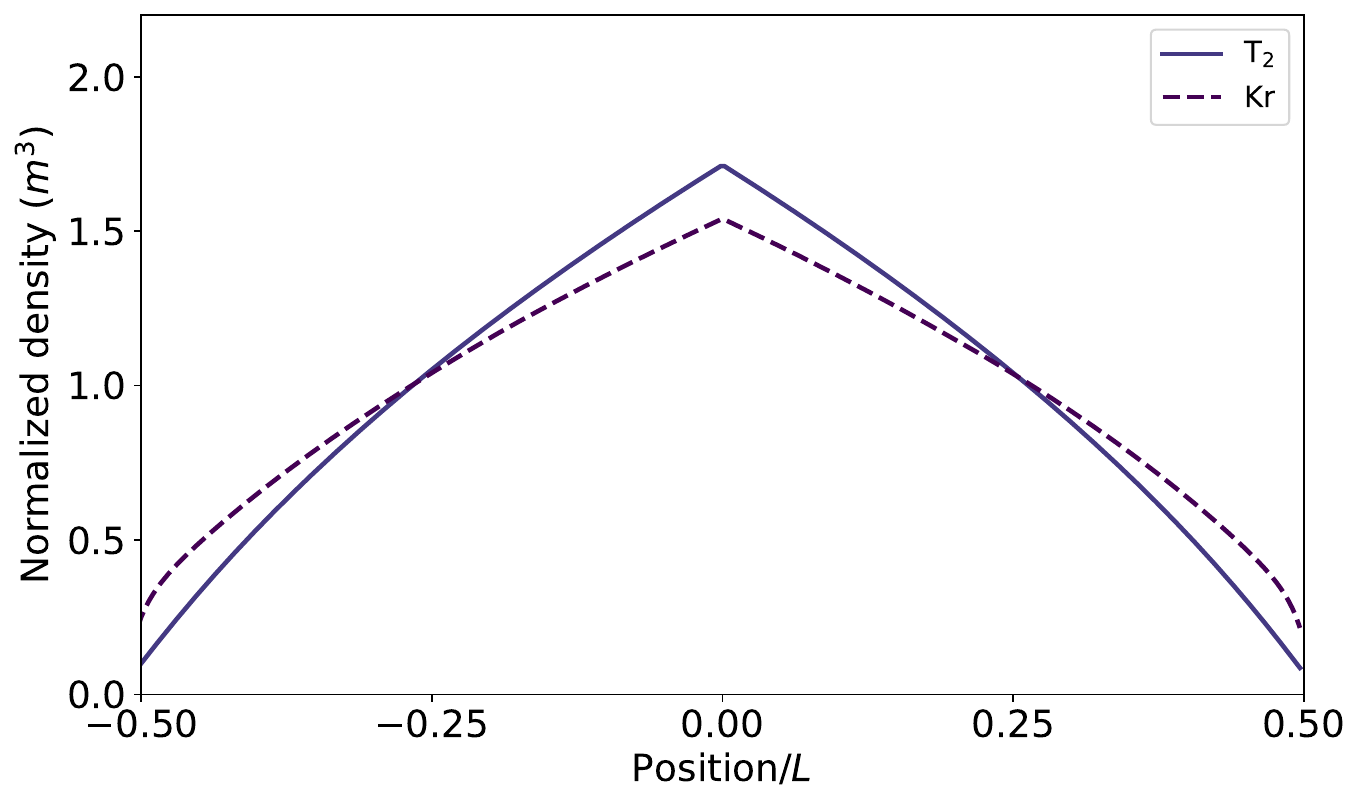}
    \caption{\textbf{Longitudinal density profile in the source.}
    The gas gets injected in the center and evacuated towards both ends of the source. Here, the densities are normalized, while the density of $\kr$ is usually around 9 orders of magnitude smaller than that of $\tr$~\cite{Mac21}. Due to the larger mass of $\kr$ it accumulates at both ends of the source compared to $\tr$. The densities were calculated following the models explained in~\cite{kalempa2010separation,Hoetzel}.}
    \label{fig:gasprofiles}
\end{figure}

The probability $p_i(E,\theta,z)$ of scattering $i$ times is given by~\cite{kleesiek2019upbeta}
\begin{equation}
    p_i(E,\theta,z)=\frac{(\mathcal{N}_\mathrm{eff}(\theta,z)\sigma_\mathrm{tot}(E))^i}{i!}\mathrm{e}^{-\mathcal{N}_\mathrm{eff}(\theta,z)\sigma_\mathrm{tot}(E)}~. \label{eq:ScatteringProbabilities}
\end{equation}
It depends on the electron energy via the scattering cross section $\sigma_\mathrm{tot}(E)$, and on the effective gas column density~\cite{kleesiek2019upbeta}
\begin{equation}
    \mathcal{N}_\mathrm{eff}(\theta,z)=\frac{1}{\cos{\theta}} \int_z^{L/2} \rho(z') \mathrm{d}z'
\end{equation}
an electron needs to traverse, where $\rho(z)$ is the density of the gas. Both a higher angle $\theta$ and a longitudinal starting position $z$ towards the rear of the source lead to a larger effective path length and thus to a higher probability of scattering.

The source strength at KATRIN is usually quantified using the column density $\mathcal{N}=\mathcal{N}_\mathrm{eff}(\theta=0,z=-L/2)$. To account for the longitudinal distribution of the electron emitters, the normalized density $\Bar{\rho}(z)=\frac{\rho(z)}{\mathcal{N}/L}$ is used. As shown in figure~\ref{fig:gasprofiles}, since the gas is injected at the center of the source and evacuated at both ends, the profile of the density is approximately triangular. For the measurements discussed in this work, gas mixtures of $\tr$ and $\kr$ are used, where the density of gaseous $\kr$ is many orders of magnitude lower than that of $\tr$. Thus, while the rate of the $\tr$ and $\kr$ spectra are proportional to their respective densities, the scattering probabilities are always calculated using the $\tr$ density.

As stated initially, the spectrum measured in a single pixel is proportional to the total number of electrons emitted along a magnetic field line. To perform the corresponding integration, all $z$-dependent quantities must be taken into account. In this section, only the scattering probabilities and the gas density are considered, while the influence of other effects is discussed in section~\ref{sec:HigherOrders}. In the absence of an electric potential, the integration over $z$ yields the weighted scattering probabilities
\begin{equation}
    \Bar{p}_i(E,\theta)\equiv\Braket{\Bar{\rho}(z) p_i(E,\theta,z)}~,
    \label{eq:WeightedScatProbs}
\end{equation}
which appear as an overall factor in the rate calculation.
\begin{figure}
    \centering
    \includegraphics[width=.49\textwidth]{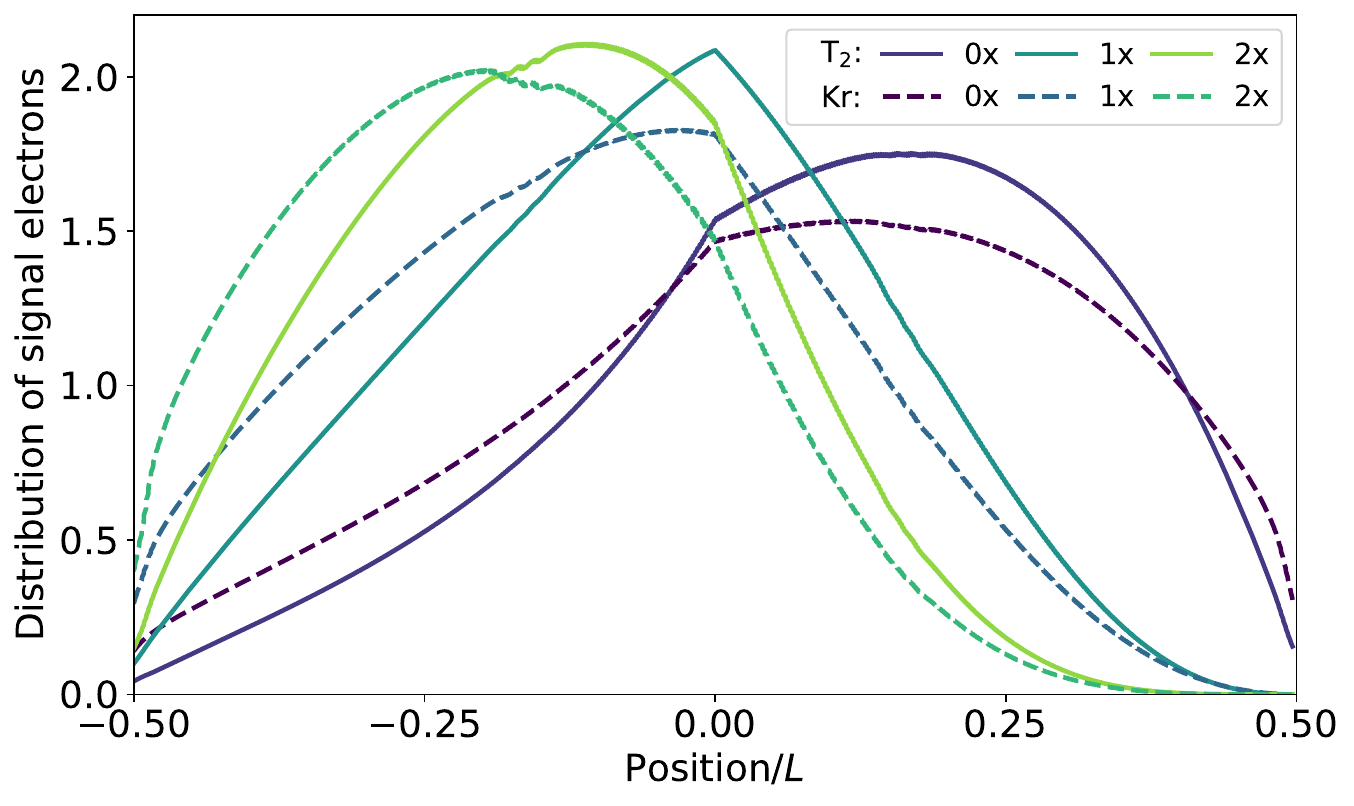}
    \caption{\textbf{Distribution of electrons in the gaseous source.}
    The electron distributions $P_i(z)$ are dominated by the $z$-dependent scattering probabilities (multiplicity 0x, 1x, 2x) and the gas density. A small, additional effect stems from a slight decrease of the source magnetic field due to gaps between the coils, causing the ripples visible around $\pm 0.2~L$ (c.f.\ section~\ref{sec:HigherOrders}). The source conditions of the KATRIN measurement campaigns since KNM3-NAP (c.f.\ table~\ref{tab:ElossCorrelations}) have been used. The magnetic field and scattering probabilities were calculated following the models explained in~\cite{Hoetzel,kleesiek2019upbeta}.}
    \label{fig:ElectronDistributions}
\end{figure}

To describe the transmission probability of an electron of energy $E$, given the retarding potential $U$, the source scattering and transmission properties of the MAC-E filter are then combined into the so-called response function
\begin{equation}
    R(E,U)=\sum_i R_i(E,U)~,
\end{equation}
where the response function for each scattering (c.f.\ equations~20 and 24 in~\cite{kleesiek2019upbeta})~\footnote{To save on space, in some cases the arguments $(E,\theta,U)$ are put at the end of the equation. The actual dependencies should be clear from the text.}
\begin{equation}
    R_i(E,U)\equiv\int\displaylimits_{\theta=0}^{\theta_\mathrm{max}}\mathrm{d}\theta\sin{\theta}\left(\Bar{p}_i\mathcal{T}\otimes f_i\right)(E,\theta,U)
    \label{eq:response}
\end{equation}
is understood as follows. $\tr$ and $\kr$ are isotropical electron sources. Therefore, the number of electrons with starting angle $\theta$ is proportional to $\sin{\theta}\Bar{p}_i(E,\theta)$. The electrons are transmitted and counted in a single pixel if the transmission condition $\mathcal{T}(E,\theta,U)$ is fulfilled and if $\theta$ does not exceed the maximum allowed angle $\theta_\mathrm{max}$. The spectrum of the electrons is modified by the inelastic scattering with probability density $f_i$.

Using the decay rate $\frac{\mathrm{d}\dot{N}}{\mathrm{d}E}\left(E\right)$ of one $\tr$ molecule or $\kr$ atom, the measured count rate is then calculated as
\begin{equation}
    \dot{N}(U)\propto\mathcal{N}\int_{-\infty}^\infty\mathrm{d}E\frac{\mathrm{d}\dot{N}}{\mathrm{d}E}\left(E\right)R(E,U)~.
    \label{eq:MeasuredRate}
\end{equation}
The constant of proportionality contains $z$-independent factors, that are irrelevant in the context of this work.

When the starting potential $V(z)$ is introduced, the transmission condition needs to be modified and the $z$-integration needs to include the potential, leading to the modified response function
\begin{equation}
    R'_i(E,U)=\int\displaylimits_{\theta=0}^{\theta_\mathrm{max}}\mathrm{d}\theta\sin{\theta}\left(\Bar{p}_i\Braket{P_i\mathcal{T}(U-V)}\otimes f_i\right)(E,\theta)~,
    \label{eq:responsePotential}
\end{equation}
with the normalized electron distributions
\begin{equation}
    P_i(E,\theta,z)\equiv\frac{\Bar{\rho}(z)p_i(E,\theta,z)}{\Bar{p}_i(E,\theta)}~.
    \label{eq:ElectronDistributionsFull}
\end{equation}
These distributions quantify the regions in which the measurement is sensitive to the starting potential, given the energy, angle, and scattering multiplicity of the electrons. They are shown in figure~\ref{fig:ElectronDistributions}, using scattering probabilities averaged over $\theta$. The difference between $\tr$ and $\kr$ arises not only from their different density profiles, but also from the fact that different energy ranges are used in the measurements ($\upbeta$\,spectrum endpoint $E_0\approx18.6\,\mathrm{keV}$, $\kr$ lines at $\approx 32.1\,\mathrm{keV}$). This results in a different inelastic scattering cross section. Consequently, the spectra of the two species are affected by the starting potential with different weights, leading to uncertainties discussed in section~\ref{sec:KrT2Differences}.

\subsection{Starting-potential distributions}
\label{sec:SPDs}
\begin{figure}
    \centering
    \includegraphics[width=.49\textwidth]{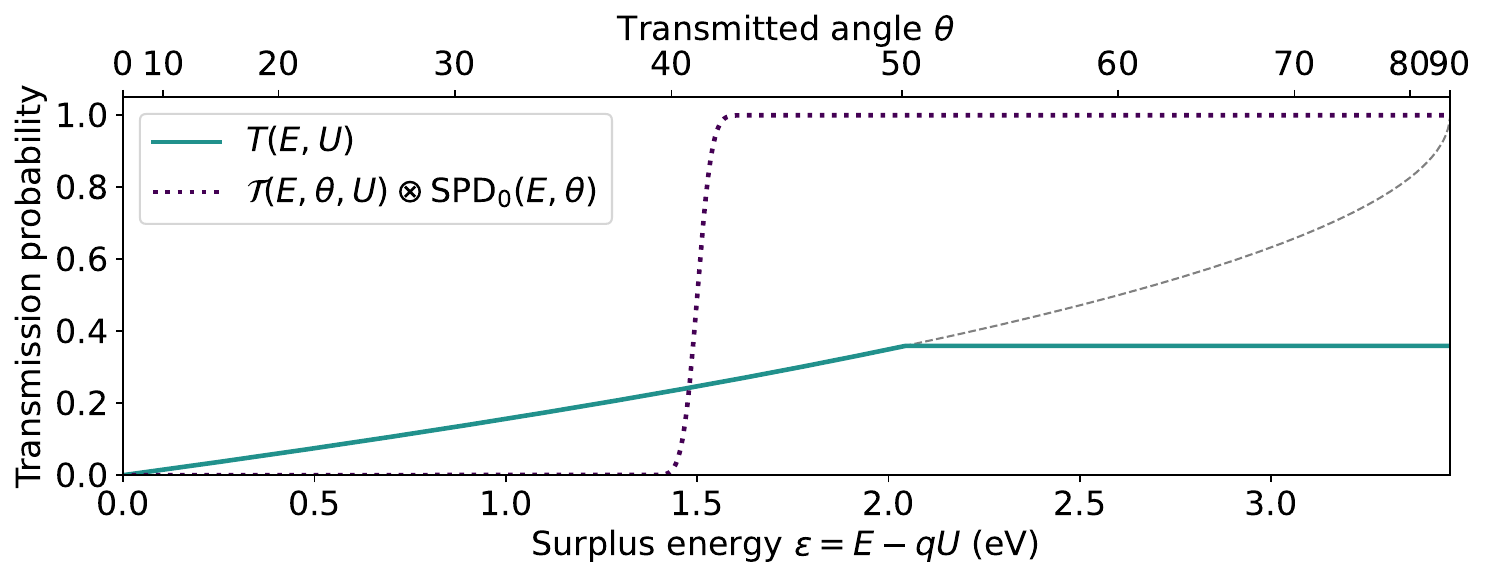}
    \caption{\textbf{Transmission function.}
    This plot illustrates equation~\ref{eq:ResponseSPDConvolution}. The transmission condition is a step function in the surplus energy, corresponding to the energy required for electrons at a given angle to be transmitted. It is modified by the potential distribution, here modeled as a Gaussian, which may also depend on the angle due to the weighting with the signal electron distributions. For unscattered electrons, the transmission function is obtained by integrating this broadened step function over the angle, taking into account the phase space factor $\sin{\theta},\Bar{p}_0(E,\theta)$. The integration is cut at the maximum angle of around $51\degree$. The plot shows typical settings used for the $\kr$ measurement.}
    \label{fig:TransmissionFunction}
\end{figure}

By inserting an integral over a $\delta$ distribution or using variable transformation, equation~\ref{eq:responsePotential} can be rewritten using an additional convolution
\begin{equation}
    R'_i(E,U)=\int\displaylimits_{\theta=0}^{\theta_\mathrm{max}}\mathrm{d}\theta\sin{\theta}\left(\Bar{p}_i\mathcal{T}\otimes\mathrm{SPD}_i\otimes f_i\right)(E,\theta, U)
    \label{eq:ResponseSPDConvolution}
\end{equation}
with normalized starting-potential distributions
\begin{equation}
    \mathrm{SPD}_i(\nu,E,\theta)\equiv\Braket{\delta(\nu-V(z))P_i(E,\theta,z)}~.
\end{equation}
They represent the frequency distributions of encountering a potential value $\nu$, considering the electron distributions $P_i(E,\theta,z)$.

Equation~\ref{eq:ResponseSPDConvolution} is visualized in figure~\ref{fig:TransmissionFunction}. It has important implications: The effect of the starting potential is fully described by the starting-potential distributions. Inversely, different starting potentials that lead to the same distributions are non-distinguishable in measured electron spectra. Any measurement of the electron spectrum can only depend on the distributions, but not the $z$-dependent potential.

In the following, it is assumed that the dependency of the starting potential distributions on the energy and the angle can be neglected:
\begin{equation}
    \mathrm{SPD}_i(\nu)\approx\Braket{\delta(\nu-V(z))P_i(z)}~,
    \label{eq:SPD}
\end{equation}
using the electron distributions approximated as
\begin{equation}
    P_i(z)\approx\frac{\Bar{\rho}(z)p_i(z)}{\Bar{p}_i}~,
\end{equation}
where the $p_i(z)$ are the scattering probabilities averaged over the angle. This approximation only concerns the sensitive region to the starting potential. It is not applied to the weighted scattering probabilities $\Bar{p}_i(E,\theta)$ of equation~\ref{eq:ResponseSPDConvolution}, which are directly proportional to the electron rate. The implications are further discussed in section~\ref{sec:HigherOrders}.

Since the convolution is both commutative and associative, the starting potential can then be considered as a modification of the energy-loss functions in equation~\ref{eq:elossitimes}
\begin{equation}
    f_i^{'}=\mathrm{SPD}_i \underbrace{\otimes f \otimes f ...}_{i-\mathrm{times}}~.
    \label{eq:SPDConvolution}
\end{equation}
Again, this has important implications: For a given scattering multiplicity, the effect of the starting potential is indistinguishable from a modification of the observed energy loss. This is a consequence of the $z$-dependency of the scattering probabilities, such that electrons with different scattering multiplicities start in different average potentials.

Thus, implementing the SPD$_i$ like in equation~\ref{eq:SPDConvolution} is the only change of the KATRIN model necessary to consider the starting potential. In particular, this approach is still correct within the discussed approximations, even if higher order effects for example of the transmission condition are taken into account.

When performing the integration over $\theta$, one obtains the transmission function~\cite{Groh,kleesiek2019upbeta}
\begin{equation}
    T_i(E,U)=\int\displaylimits_{\theta=0}^{\theta_\mathrm{max}}\mathrm{d}\theta\sin{\theta}\Bar{p}_i(E,\theta)\mathcal{T}(E,\theta,U)
\end{equation}
for each scattering $i$, which is a step function with a width $\Delta E$ on the order of 1\,eV. The spectrum is then calculated using the integration of equation~\ref{eq:MeasuredRate} and the modified response function
\begin{equation}
    R'(E,U)=\sum_i T_i(E,U)\otimes\mathrm f'_i~.
\end{equation}

\subsection{Statistical moments of the starting-potential distributions}
\label{sec:StatisticalMoments}

To quantify the effect of the starting potential on the spectra, the spectral shape has to be taken into account. Due to the commutativity of the convolution, it is also possible to first perform the energy integration, and afterwards the convolution with the SPD$_i$. This leads to the measured spectrum
\begin{equation}
    \dot{N}(U)=\sum_i \dot{N}_i(U)\otimes\mathrm{SPD}_i~,
    \label{eq:SPDSpectrumSummation}
\end{equation}
with the definition of the $i$-times scattered spectrum
\begin{equation}
    \dot{N}_i(U)\propto\mathcal{N}\int_{-\infty}^\infty\mathrm{d}E\frac{\mathrm{d}\dot{N}}{\mathrm{d}E}\left(E\right)R_i(E,U)~.
\end{equation}
Experimentally, the full SPD$_i$ cannot be determined and, instead, they are approximated by their statistical moments
\begin{equation}
    \mu_{n,i}[V]\equiv\Braket{V^n(z) P_i(z)}~.
\end{equation}
The moments then appear in the calculation of equation~\ref{eq:SPDSpectrumSummation} by means of the Taylor expansion
\begin{align}
    \dot{N_i}(U)\otimes\mathrm{SPD}_i&=\int\mathrm{d}\nu\,\dot{N_i}(U-\nu)\mathrm{SPD}_i(\nu)~, \\
    &=\sum_{n=0}^\infty \frac{\dot{N}_i^{(n)}(U)}{n!}(-1)^n\int\mathrm{d}\nu\,\nu^n\mathrm{SPD}_i(\nu)~, \\
    &=\sum_{n=0}^\infty \frac{\dot{N}_i^{(n)}(U)}{n!} (-1)^n\mu_{n,i}[V]~.
    \label{eq:MomentTaylor}
\end{align}
Here, $\dot{N}_i^{(n)}$ is the $n$-th derivative of the $i$-times scattered spectrum with respect to $U$.

If the derivatives of the spectrum decrease sufficiently fast, the higher-order terms of the expansion can be neglected or treated as biases on the leading order moments. For the tritium spectrum, Slezak argues in~\cite{Sle16}, that by this logic a second order expansion is sufficient to characterize the neutrino-mass bias, which is further discussed in section~\ref{ch:TritiumSystematics}. Accordingly, in order to characterize the potential-induced neutrino-mass bias, it is enough to determine the first two moments in the $\kr$ measurement, i.e.\ the means
\begin{equation}
    \braket{V}_i\equiv\int \frac{\mathrm{d}z}{L} V(z)P_i(z)
\end{equation}
and standard deviations
\begin{equation}
    \sigma_i[V]\equiv\sqrt{\braket{V^2}_i-\braket{V}_i^2}~.
\end{equation}
If the higher moments are neglected, this corresponds to an approximation of the SPD$_i$ with Gaussian distributions $G(\Braket{V}_i,\sigma_i[V])$. It depends on the remainder term of equation~\ref{eq:MomentTaylor} whether neglecting higher moments of the true potential causes a sizable bias on the first two moments in the $\kr$ measurement. Using Hölder's inequality on the unknown higher moments one can show, that in general the orders $n\geq3$ are suppressed with $\left(\frac{\sigma_i}{w}\right)^n$, where $w$ is the characteristic width of the spectrum. For a standard deviation of the potential of around 30\,mV~\cite{KryptonMeasurementPaper} and the transmission function width $w\approx\Delta E$ on the 1\,eV scale in the $\kr$ measurement, each subsequent order is suppressed by a factor on the order of $10^{-2}$. Further studies are recommended to show, if this suppresses a possible bias of the first two moments sufficiently.

Since a constant potential only shifts the spectra, it is convenient to use $\braket{V}_0$ to describe the overall scale of the potential, and to replace the higher $\braket{V}_{i>0}$ by the so-called energy-loss shifts
\begin{align}
    \Delta_{i0}[V]&\equiv\braket{V}_i-\braket{V}_0~, \\
    &=\int \frac{\mathrm{d}z}{L} V(z)\left[P_i(z)-P_0(z)\right]~.
\end{align}
They are the potential induced shift of the $i$-times scattered spectra compared to the unscattered spectrum. As they vanish for a constant potential, they are measures of the inhomogeneity of the potential.

Describing the influence of the potential by the statistical moments $\braket{V}_0$, $\Delta_{i0}[V]$, $\sigma_i[V]$ is the natural, perturbative approach up to the second power of the potential. Up to this order, any observable of the potential has to depend on combinations of these moments. Also, since the energy-loss shifts are a linear order effect, their influence on the measured spectra is in general expected to be larger than that of the standard deviations. The $\Delta_{i0}$ exist as a consequence of the electron scattering on the source gas, and were not considered by KATRIN's predecessor experiments.

\subsection{Determining the moments: The gaseous $\kr$ measurement}
\begin{figure}[tp]
    \centering
    \includegraphics[width=.5\textwidth]{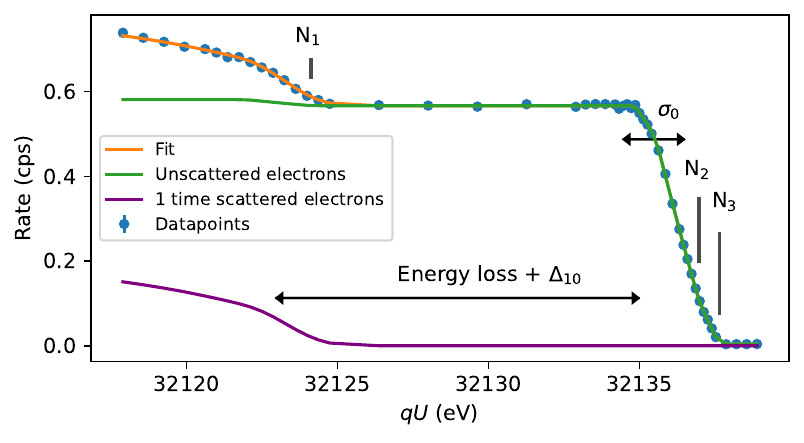}
    \caption{\textbf{Krypton-83m N$_{23}$ spectrum with energy loss.}
    The orange line shows a fit of the measured N$_{123}$ spectrum. One-time scattered electrons are visible as a rate increase approximately 13\,eV below the main peak (purple). Changes of this distance $\Delta_{10}[V]$ are an observable of the inhomogeneity of the electric potential. Changes of the line width of the non-scattered electrons are given by the standard deviation $\sigma_0[V]$ of the potential. The N$_1$ line coincides with the scattering and is a nuisance in these measurements. Plot from~\cite{KryptonMeasurementPaper}.}
    \label{fig:KryptonN123Spectrum}
\end{figure}

At KATRIN gaseous krypton-83m is co-circulated alongside tritium for the purpose of calibration measurements~\cite{Mac21,KryptonMeasurementPaper}. The krypton-83m generates a spectrum of quasi-mono-energetic conversion electrons, which can be used to determine $\braket{V}_0,\Delta_{10}[V]$ and $\sigma_0^2[V]$. The experimental details and considered sources of error are discussed in~\cite{KryptonMeasurementPaper}.

A spectrum of the measurements is found in figure~\ref{fig:KryptonN123Spectrum}. Since the width of the used N$_{23}$ lines can be approximated as zero, the spectrum is essentially a direct measurement of the response function as modified by the starting potential. Electrons lose at least around 13\,eV when scattering on the tritium gas, such that in the experimental $\kr$ spectrum the first scattering is directly visible as a line 13\,eV below the non-scattered line. Since the scattering introduces a broadening to the scattered spectrum that is orders of magnitude larger than the variance of the source potential, the broadenings $\sigma_{i\geq1}$ are irrelevant both for the krypton and tritium measurement. In the fit of the $\kr$ spectra usually $\sigma_0=\sigma_1$ is used, and the measurement range is cut such that no significant contribution from scatterings larger than 1 is observed, as shown in the plot.

The $\kr$ measurement only allows to determine $\Delta_{10}[V]$, but no energy-loss shifts for higher scatterings. Fortunately, since both the $\Delta_{i0}[V]$ and the $\sigma_i[V]$ are a measure of inhomogeneity of the potential, the allowed parameter space of all the $\Delta_{i0}[V]$ is already constrained by one measured $\sigma_j[V]$. This is shown in the following.

\subsection{Relation of the moments to the potential shape}
\label{sec:PotentialShape}
\begin{figure}[tp]
    \centering
    \includegraphics[width=.5\textwidth]{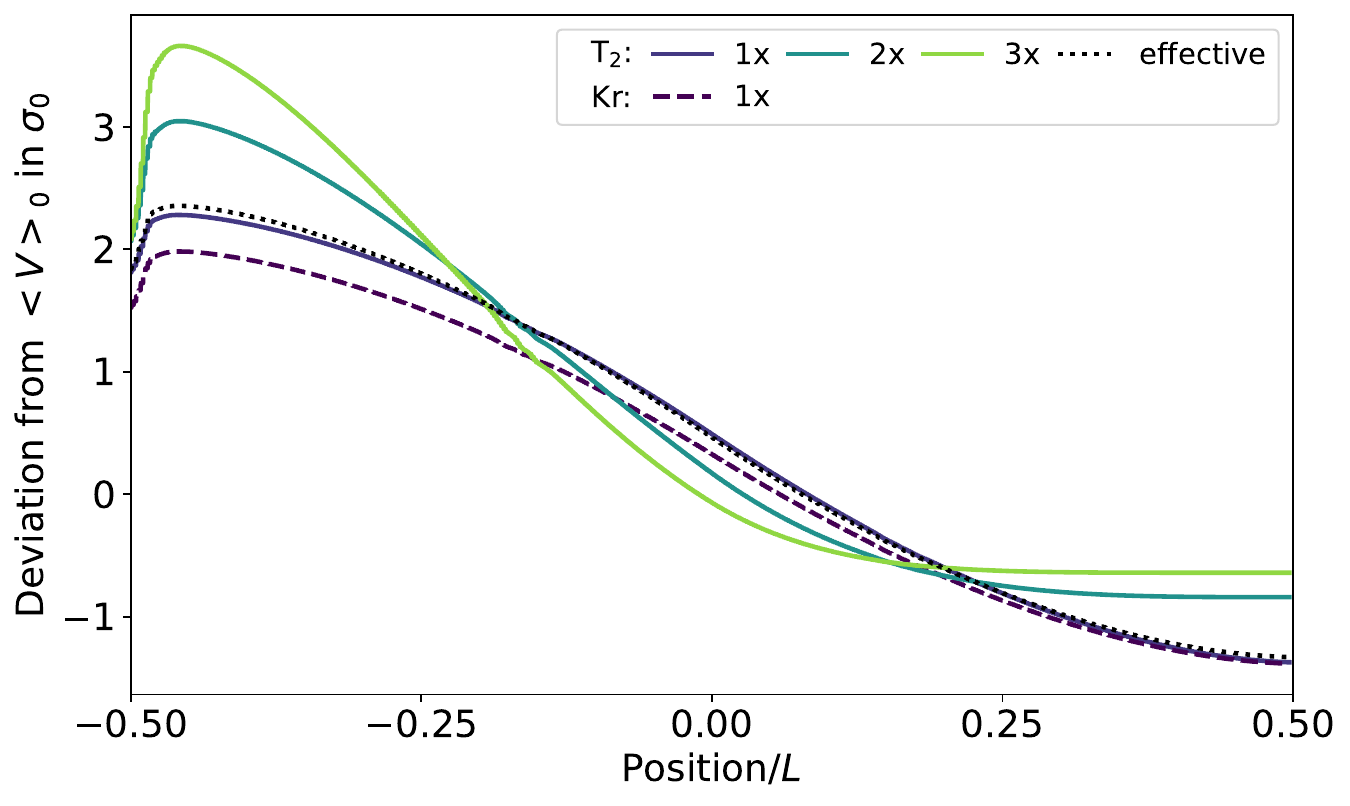}
    \caption{\textbf{Maximum solution.}
    Shown are the potential shapes $Q_i(z)$ that lead to the maximum energy-loss shifts for the scattering multiplicities 1x, 2x and 3x for tritium electrons, and 1x for krypton electrons. The "effective" shape combines the contributions from all scatterings, considering their relevance for the tritium $\upbeta$\,spectrum (c.f.\ section~\ref{sec:EffectiveDelta}). In the 40\,eV analysis window of the $\upbeta$\,spectrum, the effective shape is strongly dominated by 1x scattering. The source conditions of the KATRIN measurement campaigns since KNM3-NAP (c.f.\ table~\ref{tab:ElossCorrelations}) have been used.}
    \label{fig:Qi}
\end{figure}

In this section, relations between the observables $\Delta_{i0}$ and $\sigma_0$ are derived. This section generalizes findings from \cite{Mac21} up to arbitrary numbers of scattering $i$. The standard deviation $\sigma_0$ of the non-scattered electron distribution is used here for brevity, but any other one works the same way. This and more generalizations are discussed in~\cite{Mac21} alongside other constraints of for example the peak-to-peak value of the potential by the measured observables.

Using the weighted covariance
\begin{equation}
    \mathrm{Cov}_k[F,G]=\Braket{FG}_k-\Braket{F}_k\Braket{G}_k~,
\end{equation}
one can write the energy-loss shifts as
\begin{equation}
    \Delta_{i0}[V]=\mathrm{Cov}_0\left[\frac{P_i-P_0}{P_0},V\right]~.
    \label{eq:DeltaCovariance}
\end{equation}
This is shown using $\Braket{F}_0=\Braket{FP_0}$ and the fact that $\braket{P_i-P_0}=0$ due to the normalization of the $P_i(z)$. Thus, the second term vanishes when one expands the covariance in equation~\ref{eq:DeltaCovariance}. The goal is now to derive a measure of correlation from equation~\ref{eq:DeltaCovariance}. First, the scalar covariances of the electron distributions are introduced as
\begin{equation}
    \kappa_{ij}\equiv\mathrm{Cov}_0\left[\frac{P_i-P_0}{P_0},\frac{P_j-P_0}{P_0}\right]~,
\end{equation}
with $\kappa_{ii}\equiv\kappa_i^2$.
\begin{figure*}
    \centering
    \begin{minipage}{0.48\textwidth}
        \centering
        \includegraphics[width=\linewidth]{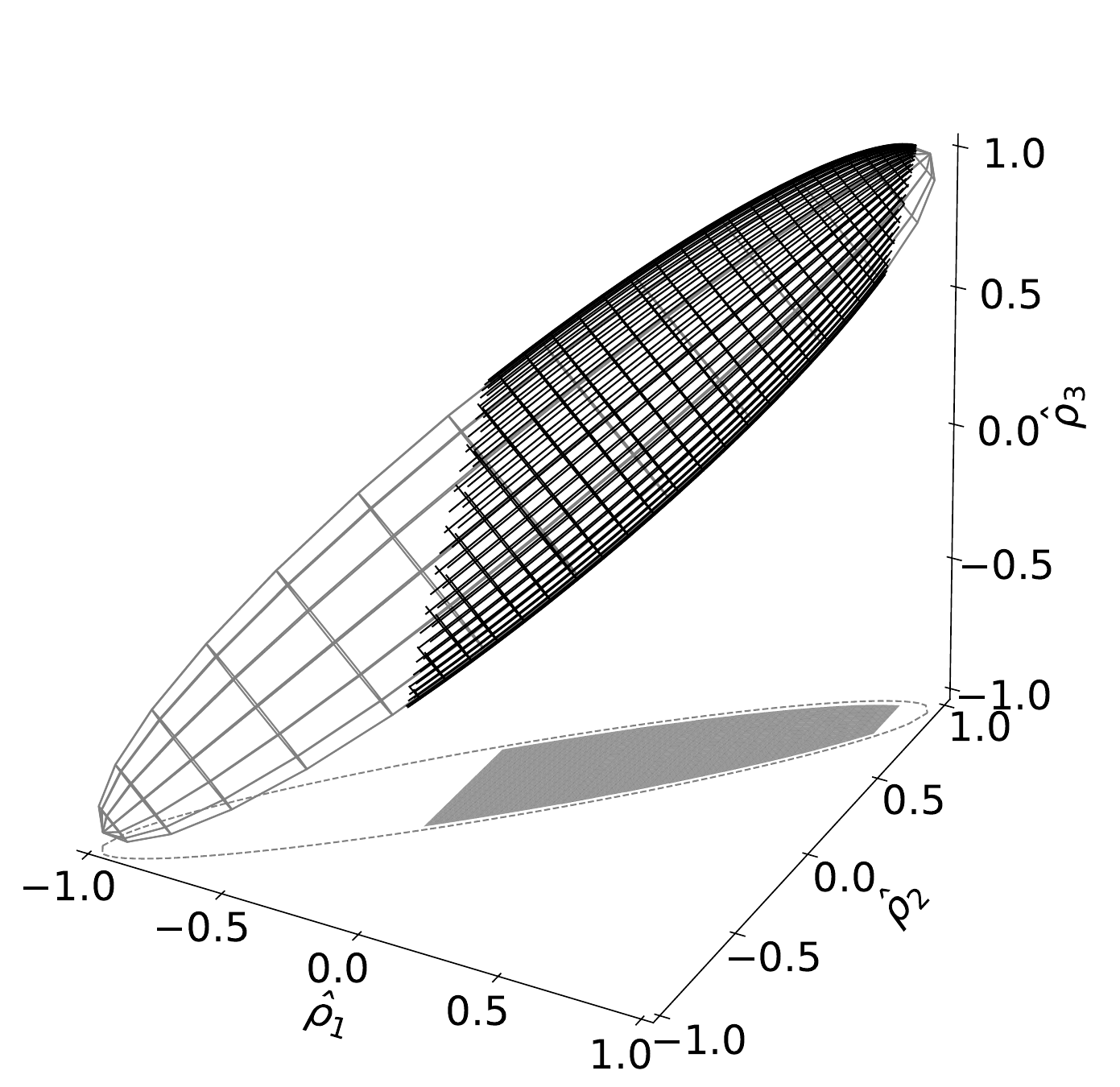}
    \end{minipage} \hfill
    \begin{minipage}{0.48\textwidth}
        \centering
        \includegraphics[width=\linewidth]{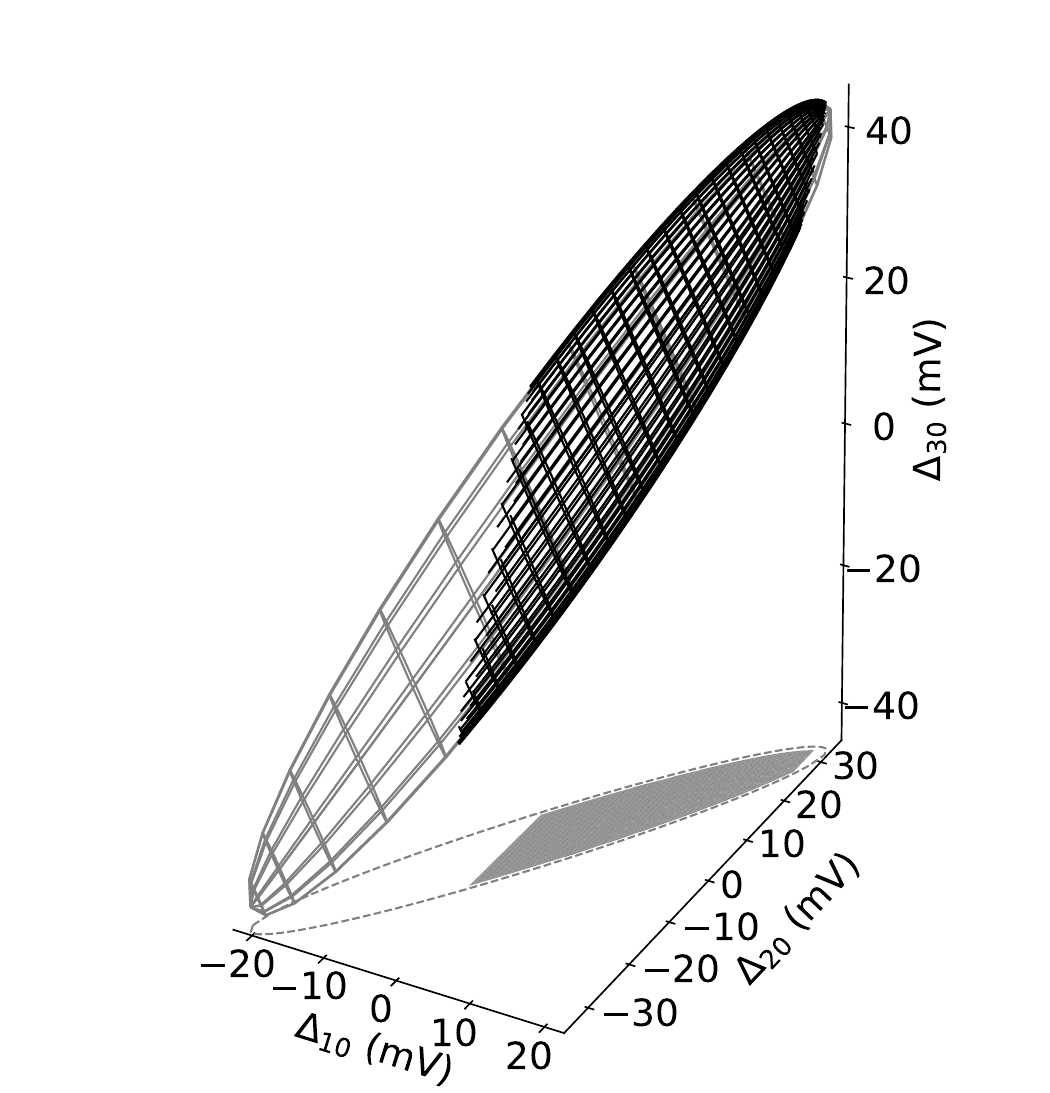}
    \end{minipage}
    \caption{\textbf{Parameter spaces.} The left plot shows the high correlation of the $\vec{\rho}$ parameter space: the major axis of the ellipsoid is oriented almost perfectly along the diagonal, such that high (low) values of one $\hat{\rho}_i$ correspond to high (low) values of the other $\hat{\rho}_j$. In the right plot the ellipsoid is rescaled to represent the $\vec{\Delta}$ parameter space, constrained by a measured value of $\sigma_0\approx\,32\,\text{mV}$~\cite{KryptonMeasurementPaper}. The constraint weakens for higher scatterings, since the overlap of the respective electron distribution with the one of the unscattered electrons decreases. Shaded in black, the $1\,\sigma$ region constrained by a $\kr$ measurement of $\Delta_{10}=(11\pm7)\,\text{mV}$ is shown, where the central value was chosen arbitrarily. The uncertainty is motivated by the already recorded experimental data. The constraint considers the difference between the $\tr$ and $\kr$ $\Delta_{10}$, as discussed in section~\ref{sec:KrT2Differences}.}
    \label{fig:ParameterSpace}
\end{figure*}

By normalizing equation~\ref{eq:DeltaCovariance} by the standard deviation of the arguments on the right side, one obtains the shape operators
\begin{equation}
    \hat{\rho}_i[V]\equiv\frac{\mathrm{Cov}_0\left[\frac{P_i-P_0}{P_0},V\right]}{\sigma_0\left[\frac{P_i-P_0}{P_0}\right]\sigma_0[V]}=\frac{1}{\kappa_i}\frac{\Delta_{i0}}{\sigma_0}[V]~.
    \label{eq:ShapeOperator}
\end{equation}
Since the left side defines a correlation, which is bounded in the interval $[-1,1]$, it holds for any arbitrary potential that
\begin{equation}
    \left|\Delta_{i0}[V]\right|\leq\kappa_i\sigma_0[V]~.
    \label{eq:Inequality}
\end{equation}
The equal sign holds for the potentials
\begin{equation}
    \frac{V(z)-\braket{V}_0}{\sigma_0}=\pm\underbrace{\frac{1}{\kappa_i}\frac{P_i-P_0}{P_0}(z)}_{\equiv Q_i(z)}~,
    \label{eq:ExtremalPotential1D}
\end{equation}
which directly follows from equation~\ref{eq:ShapeOperator}, using that the covariance of the constant term $\braket{V}_0$ vanishes. The $\hat{\rho_i}$ measure the correlation to this potential shape $Q_i(z)$, and thus the absolute value of $\Delta_{i0}$ is larger on the scale of $\sigma_0$, when the potential is similar to equation~\ref{eq:ExtremalPotential1D}.

A plot of the $Q_i(z)$ is shown in figure~\ref{fig:Qi}. Qualitatively, non-scattered electrons stem from the front side of the WGTS, while scattered electrons stem from the rear side, such that antisymmetric potentials with respect to the central injection point maximize $|\Delta_{i0}|$ for a given $\sigma_0$. Symmetric potentials with regard to the injection point do not produce large $|\Delta_{i0}|$ compared to $\sigma_0$.

\subsubsection*{Relations for arbitrary number of scatterings}
\begin{table*}
    \centering
    \caption{Standard deviations $\kappa_i$ and correlations $\rho_{ij}$ of the weight functions for the different KATRIN measurement campaigns KNMx. The last column indicates which value of $\sigma_0$ must be used together with the coefficients. The reason for this is explained in section~\ref{sec:KrT2Differences}. The values of the column density (CD) are relative to the nominal value of $5\cdot10^{21}\frac{1}{\mathrm{m}^2}$.}
    \begin{tabular}{ccccccccc}
    \toprule
         Measurement & Configuration & $\kappa_1$ & $\kappa_2$ & $\kappa_3$ & $\rho_{12}$ & $\rho_{13}$ & $\rho_{23}$ & value of \\
         campaign &  (CD, Temp) & & & & & & & $\sigma_0$ from \\
    \midrule
        KNM1 & \SI{20}{\percent}, \SI{30}{\kelvin} & 0.62 & 0.97 & 1.23 & 0.97 & 0.91 & 0.98 & $\tr$ \\
        KNM2 & \SI{84}{\percent}, \SI{30}{\kelvin} & 0.75 & 1.23 & 1.62 & 0.96 & 0.89 & 0.98 & $\tr$ \\
        KNM3-SAP & \SI{40}{\percent}, \SI{80}{\kelvin} & 0.62 & 0.97 & 1.23 & 0.96 & 0.90 & 0.98 & $\kr$ \\
        Since KNM3-NAP & \SI{75}{\percent}, \SI{80}{\kelvin} & 0.69 & 1.09 & 1.38 & 0.96 & 0.89 & 0.98 & $\kr$ \\
    \bottomrule
    \end{tabular}
    \label{tab:ElossCorrelations}
\end{table*}

For KATRIN more than one scattering is relevant. Now, the image $\Vec{\rho}[V]=(\hat{\rho}_{1}~\hat{\rho}_{2}~\dots)^\mathrm{T}[V]$ is bounded for any $V$. Since the $\Delta_{i0}$ are linear, one obtains the boundary potentials as linear combination of the one-dimensional solutions
\begin{equation}
    \frac{V(z)-\braket{V}_0}{\sigma_0}=\vec{a}\Vec{Q}(z)~,
    \label{eq:LinearCombination}
\end{equation}
where $\vec{a}$ is the coefficient vector. Its normalization is determined by calculating $\sigma_0[\dots]$ of equation~\ref{eq:LinearCombination}. Using the scalar correlations
\begin{align}
    \rho_{ij}&\equiv\frac{\kappa_{ij}}{\kappa_i \kappa_j}~, \\
    &=\hat{\rho}_i[Q_j]~, \label{eq:RhoGeneration}
\end{align}
with corresponding correlation matrix $\korrM$, basic algebra of covariances and equation~\ref{eq:RhoGeneration}, it can be shown that
\begin{align}
    \sigma_0[\vec{a}\vec{Q}]&=\sqrt{\vec{a}^\mathrm{T}\korrM\vec{a}}~, \\
    \vec{\rho}[\vec{a}\vec{Q}]&=\frac{\korrM\vec{a}}{\sqrt{\vec{a}^\mathrm{T}\korrM\vec{a}}}~,
    \label{eq:RhoCorrelationCoefficient}
\end{align}
holds. Since $\sigma_0[\dots]$ applied to the left side of equation~\ref{eq:LinearCombination} is 1, $\vec{a}$ needs to be chosen such that
\begin{equation}
    \vec{a}^\mathrm{T}\korrM\vec{a}=1~.
    \label{eq:CoefficientVectorEllipsoid}
\end{equation}
Now, using the inverse of the correlation matrix $\korrM^{-1}$ and equations~\ref{eq:RhoCorrelationCoefficient} and~\ref{eq:CoefficientVectorEllipsoid}, the coefficient vector of the boundary potentials is determined to be
\begin{equation}
    \vec{a}=\korrM^{-1}\vec{\rho}
\end{equation}
and the boundary of $\vec{\rho}$ is given by
\begin{equation}
    1=\Vec{\rho}^\mathrm{T}\korrM^{-1}\Vec{\rho}~.
    \label{eq:RhoVecEquation}
\end{equation}
Since the correlation matrix is positive definite, this parametrizes an ellipsoid, which is shown in figure~\ref{fig:ParameterSpace}. A maximum of 3 scatterings is shown, with correlations derived from the electron distributions at the source density that was used since the KNM3-NAP neutrino-mass measurement campaign at KATRIN. Values of the covariances and correlations of the weight functions are shown in table~\ref{tab:ElossCorrelations}.

\section{Effect of the source potential on the $\tr$-spectrum observables}
\label{ch:TritiumSystematics}

The goal of this section is to derive equations for the leading order changes of the observed tritium spectrum endpoint $\Delta E_0$ and the squared neutrino mass $\Delta m^2$, caused by an unknown starting potential. The main sources on this topic \cite{RobertsonKnapp,Sle16} do not consider electron scattering and thus do not account for a possible bias caused by energy-loss shifts $\Delta_{i0}[V]$. To do so, the response function of KATRIN has to be considered, which leads to a very complex discussion. Thus, the aim of this section is to gain a qualitative understanding rather than a quantitative one, and to show, that there are significant deviations to the previous understanding of this kind of systematic bias on the neutrino mass. The quantitative assessment is best done on Asimov Monte Carlo studies on the detailed model, as in~\cite{Mac21}.

\subsection{Small energy fluctuations of the intrinsic $\upbeta$\,spectrum}

It is long established, that the leading order bias of the squared neutrino mass, caused by an unrecognized Gaussian energy distribution convolved with the tritium $\upbeta$\,spectrum, is given by $\Delta m^2=-2\sigma^2$, where $\sigma^2$ is the variance of the distribution~\cite{RobertsonKnapp}. Naturally, the endpoint is shifted by the mean $\mu$ of the distribution $\Delta E_0=\mu$. In \cite{Sle16}, Slezak showed these relations for an arbitrary, normalized energy distribution $f$ and also concluded, that the next higher moment, the skewness, has no additional effect on $\Delta E_0$ and $\Delta m^2$.

Thus, for the intrinsic $\upbeta$\,spectrum, following from~\cite{Sle16} the leading order shifts are
\begin{align}
\begin{split}
    \Delta E_0&=\mu[f]~,    \\
    -\frac{\Delta m^2}{2}&=\sigma^2[f]~,
    \label{eq:BetaShifts}
\end{split}
\end{align}
where
\begin{align}
\begin{split}
    \mu[f]&\equiv\int\mathrm{d}\epsilon\,\epsilon f(\epsilon)~,    \\
    \sigma^2[f]&\equiv\int\mathrm{d}\epsilon\,(\epsilon-\mu[f])^2 f(\epsilon)~,
    \label{eq:PDFMoments}
\end{split}
\end{align}
are the mean and central variance of the distribution $f$.
\begin{figure}
    \centering
    \includegraphics[width=.5\textwidth]{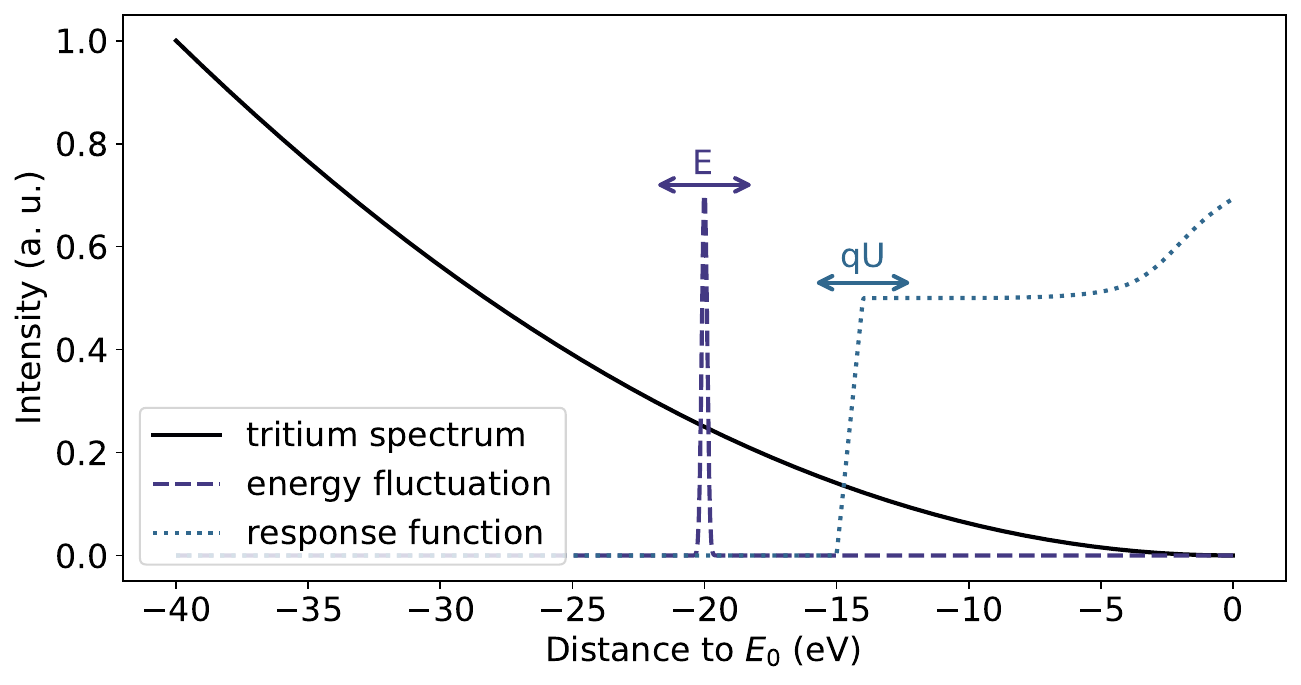}
    \caption{\textbf{Integrations over the tritium spectrum.}
    In the convolution with an energy fluctuation, the spectrum is averaged around the selected energy $E$, weighted with the fluctuation. Since the width of the fluctuation is small, the cut by the $\upbeta$\,spectrum is only relevant for energies $E$ very close to the endpoint $E_0$. In contrast, the integration over the response function depends on the upper cut of the $\upbeta$\,spectrum for every $qU$.}
    \label{fig:BetaConvolution}
\end{figure}

Slezak derives the factors $\epsilon,\epsilon^2$ in the integrals of equation~\ref{eq:PDFMoments} from the Taylor expanded $\upbeta$\,spectrum for small neutrino mass. However, as visible in figure~\ref{fig:BetaConvolution}, to be fully precise these integrals need to contain the energy cut by the endpoint $\Theta(E_0-m-E+\epsilon)$, such that the boundaries can come either from the endpoint, or from the energy distribution $f$. The transition happens at energies where $E_0-m-E\lesssim\sigma[f]$. If $\sigma[f]$ is small, for example on a sub eV scale, then this subtlety really only arises so close to the endpoint, that it is statistically insignificant given the small rate in this region and that is what \cite{Sle16} implicitly uses. However, when the response function is introduced, the cut by the endpoint becomes important, as discussed in the following.

\subsection{Systematic bias as moments of the response function}

Despite of the sign of the integration variable, the integration over the response function as in equation~\ref{eq:MeasuredRate} is the same operation as the convolution with an energy fluctuation. Ultimately, in order to derive the shift of the $\upbeta$\,spectrum observables caused by an unrecognized modification of the response function in leading order, at some point the Taylor expansion of the spectrum introduced in~\cite{Sle16} has to be used, again leading to the equations~\ref{eq:BetaShifts}, only that the energy fluctuation is replaced by the response function.

However, to do so, the limits of the integrals have to be considered. As visible in figure~\ref{fig:BetaConvolution}, since the upper cut always comes from the endpoint, it is sufficient that the response function vanishes at the lower side. One can then define the moments of the response function $R$ like
\begin{align}
\begin{split}
    N[R]&=\int^{E_0-m}\mathrm{d}E\,R(E)~,    \\
    \mu[r]&=\int^{E_0-m}\mathrm{d}E\,E\,r(E)~,    \\
    \sigma^2[r]&=\int^{E_0-m}\mathrm{d}E\,(E-\mu[r])^2\,r(E)~,
    \label{eq:PDFMomentsCut}
\end{split}
\end{align}
where $r$ is $R$ divided by the norm $N[R]$ of $R$. $N[R]$ is only a scaling factor, which can be absorbed by a nuisance parameter of the $\upbeta$\,spectrum intensity. These moments implicitly depend on $U$ via the response function.

Fitting a model using a response function $r$ without potential to data with response $r'$, that includes the potential, one then obtains the shifts
\begin{align}
\begin{split}
    \Delta E_0&=\Braket{\mu[r]-\mu[r']}_{U}~, \\
    -\frac{\Delta m^2}{2}&=\Braket{\sigma^2[r]-\sigma^2[r']}_{U}~,
    \label{eq:BetaShiftsDifferences}
\end{split}
\end{align}
as average over all $U$. Here, some major differences to the shifts obtained for the intrinsic $\upbeta$\,spectrum (equations~\ref{eq:BetaShifts}) are visible: The weighting of the $U$-average has to contain the distribution of measurement time to the retarding potentials, such that the bias does depend on the measurement conditions\footnote{Also, usually the observables are derived from a $\chi^2$ minimum, in which case the weight also contains the normalized derivative of the integrated spectrum with regard to each observable. This gives a system of equations, which modifies the results compared to equations~\ref{eq:BetaShiftsDifferences}.}. This is expected, since a modification of the response function in for example the energy loss model can only affect the measurement, if scattered electrons are part of the measurement time distribution. Also, in contrast to equations~\ref{eq:BetaShifts}, where the shift of the squared neutrino mass is always negative, in equation~\ref{eq:BetaShiftsDifferences} it can also be positive.

\subsection{Leading order calculation}
\begin{figure}
    \centering
    \includegraphics[width=.5\textwidth]{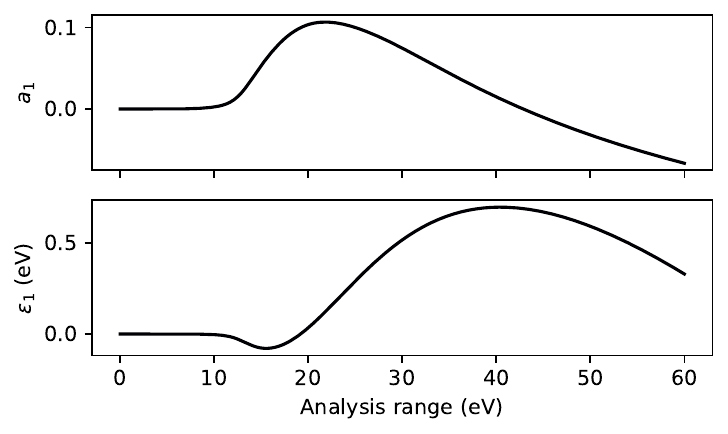}
    \caption{\textbf{Susceptibility of endpoint (top) and squared neutrino mass (bottom) to an unaccounted $\bm{\Delta}_{\mathbf{10}}$ vs.\ analysis range.}
    Shown are the first derivatives of the shifts $\Delta E_0$ and $\Delta m^2$ with regard to an unaccounted $\Delta_{10}$ (c.f.\ equations~\ref{eq:Susceptibilities}). The shifts were calculated from equations~\ref{eq:LinearSusceptibilities} using an approximate response model. Compared to the full simulation in~\cite{Mac21}, good qualitative agreement is observed for the case of $\epsilon_1$. In contrast, while the scale of $a_1$ is correct, the position of its maximum deviates visibly, possibly due the approximations of the analytical calculation.}
    \label{fig:AnalysisRange}
\end{figure}
Evaluating equations~\ref{eq:BetaShiftsDifferences} analytically is very challenging. To simplify the process they are linearized. Thus, the goal is to derive the leading order coefficients
\begin{align}
\begin{split}
    a_i\equiv&\frac{1}{q}\frac{\mathrm{d}E_0}{\mathrm{d}\Delta_{i0}}~, \\
    \epsilon_i\equiv&\frac{1}{q}\frac{\mathrm{d}m^2}{\mathrm{d}\Delta_{i0}}~,
    \label{eq:Susceptibilities}
\end{split}
\end{align}
i.e.\ the first derivative of the endpoint and squared neutrino mass with regard to $\Delta_{i0}$. Since the calculation is very technical and the results are only qualitative, only the basic ideas are described here. $q=-1$ is used in the following.

\paragraph{Normalization:} The upper integration boundary of the response moments (equations~\ref{eq:PDFMomentsCut}) depends on the bias of the endpoint and the squared neutrino mass. It is assumed, that the dependency on the neutrino mass is covered by the background. To compensate for the remaining dependency, a correction term proportional to the endpoint is added to the final results. It is calibrated with the case of a constant potential $V(z)=\Braket{V}_0$, where the endpoint bias is just the potential $\Delta E_0=q\braket{V}_0$ and the neutrino mass bias vanishes.
 
\paragraph{Energy dependency:} It is assumed that the response function only depends on the surplus energy $\epsilon\equiv E-qU$. In this case all integrals can be rewritten as function of the surplus energy and the distance $\Delta qU=E_0-m-qU$ of the retarding energy to the endpoint. The integrals of equation~\ref{eq:PDFMomentsCut} can then be reformulated in the range of $[0,\Delta qU]$. Thereby one finds that the mean $\mu[r]$ scales linearly with $qU$, and it is convenient to replace it by
\begin{equation}
    \Tilde{\mu}[r]\equiv \mu[r]-qU~.
\end{equation}

\paragraph{Linear combination:} Since the response function is a summation of response functions of different scatterings $R_i$, the same holds for the shifts $\Delta E_0$ and $\Delta m^2$, which then depend on moments of the normalized response like $N_i\equiv N[r_i]$, $\Tilde{\mu}_i\equiv\Tilde{\mu}[r_i]$ etc.\

\paragraph{Expansion:} The modified response functions for each scattering are expanded in orders of the potential
\begin{equation}
    \Braket{R_i(E,U-V(z))P_i(z)}=\sum_{n=0}^\infty\frac{R_i^{(n)}(E,U)}{n!}(-1)^n\braket{V^n}_i~.
    \label{eq:ResponseExpansion}
\end{equation}
$R_i^{(n)}$ is the $n$-th derivative with regard to $U$. The change of the normalized response with $\Delta_{i0}[V]$ is then
\begin{align}
    \frac{\mathrm{d}r}{\mathrm{d}\Delta_{i0}}&=\frac{R_i^{(1)}}{N[R]}-r\frac{N[R_i^{(1)}]}{N[R]}~, \\
    &\equiv r_i^{(1)}-rN[r_i^{(1)}]~,
    \label{eq:DerivativeNormalizedResponse}
\end{align}
where it is important to consider, that both the response and the normalization depend on $\Delta_{i0}$. The normalization of the first derivative $N[r_i^{(1)}]$ is just the response $r_i(\Delta qU)$ itself. Using integration by parts on some of the integrals, combining all the steps and omitting the arguments $[r]$ and $[r_i]$, one then finds the leading order shifts
\begin{align}
\begin{split}
    \Delta E_0^{(1)}&=-\braket{V}_0-\sum_{i=1}a_i\Delta_{i0}[V]~, \\
    \Delta {m^2}^{(1)}&=-\sum_{i=1}\epsilon_i\Delta_{i0}[V]~,
\end{split}
\end{align}
where
\begin{align}
\begin{split}
    a_i&=\frac{\Braket{N_i-\mathcal{E}r_i}_{\Delta qU}}{\Braket{1-\mathcal{E}r}_{\Delta qU}}~, \\
    \epsilon_i&=\Braket{4\mathcal{E}N_i-4\mathcal{E}_i+2\mathcal{V}(a_ir-r_i)}_{\Delta qU}~.
    \label{eq:LinearSusceptibilities}
\end{split}
\end{align}
The new quantities are defined and discussed in the following. The asymptotic behavior of the normalized response $r_i$ and the normalization $N_i$ is related to the weighted scattering probabilities $\Bar{p}_i$ (equation~\ref{eq:WeightedScatProbs})
\begin{align}
    \lim_{\Delta qU\to\infty}r_i&\propto \frac{\Bar{p}_i}{\Delta qU}~, \\
    \lim_{\Delta qU\to\infty}N_i&\propto \Bar{p}_i~.
\end{align}
The normalization $N_i$ can be interpreted as the fraction of measured surplus energy of $i$-times scattered electrons.

Both coefficients in equation~\ref{eq:LinearSusceptibilities} depend on a measure of the mean measured surplus energy of $i$-times scattered electrons
\begin{equation}
    \mathcal{E}_i\equiv N_i\Delta qU-\Tilde{\mu}_i~,\quad\lim_{\Delta qU\to\infty} \mathcal{E}_i\propto\Delta qU\Bar{p}_i~.
\end{equation}
Additionally, $\epsilon_i$ depends on a measure of the variance of the measured surplus energy
\begin{equation}
    \mathcal{V}\equiv\mathcal{E}^2-\sigma^2~,\quad\lim_{\Delta qU\to\infty} \mathcal{V}\propto\Delta qU^2~.
\end{equation}
An evaluation of equations~\ref{eq:LinearSusceptibilities} using a simplified response model and an unweighted $\Delta qU$-averaging is shown in figure~\ref{fig:AnalysisRange} as function of the analysis range of the $\upbeta$\,spectrum. Both graphs show a non-trivial structure. Comparing to~\cite{Mac21}, where the same coefficients were studied with fits on Asimov spectra, the order of magnitude is in agreement and the position of the visible maximum is agreeing for $\epsilon_1$. In contrast, the position of the maximum of $a_1$ does not agree with the simulation, possibly due to approximations of the calculation shown here.

\subsection{Interpretation}

Since the response function is not a density function, the meaning of the moments and the derived expressions~\ref{eq:LinearSusceptibilities} is not intuitive. Nevertheless, these equations are still useful to understand figure~\ref{fig:AnalysisRange}.

As it should be expected, $a_i$ and $\epsilon_i$ are only non-vanishing, if the respective scattering multiplicity is contained in the measurement time distribution.

The averaging over $\Delta qU$ entails an integration, such that the expressions inside the averages are proportional to the derivatives of figure~\ref{fig:AnalysisRange}. Asymptotically, the expressions inside the averages scale with $\Bar{p}_i$ in case of the endpoint and $\Delta qU \Bar{p}_i$ in case of the squared neutrino mass. However, since the response function varies on a scale of several 10\,eV, the asymptotic behavior is not reached in the plotted range, such that the curves do depend on the details of the response model. Also, the integration of the $\Delta qU$-average raises the exponent of $\Delta qU$ by one. In the exemplary case of $a_i$ this means, that the dependency on the analysis range does not vanish even for large $\Delta qU$, and the $a_i$ do not converge to $\Bar{p_i}$, as it might be naively assumed. This is a result of the integral measurement at KATRIN, since a data point at $qU$ depends on the whole spectrum for energies larger than $qU$.

Due to the complexity, the second order terms have not been investigated analytically. However, since the energy-loss function has a width that is orders of magnitudes larger than the standard deviations of the potential, only $\sigma_0$ is relevant, which was also found in the simulations of~\cite{Mac21}.

Overall, the linear order shifts of $E_0$ and $m^2$ occur because of a wrongly predicted mean and variance of the measured surplus energy. The order of magnitude of the coefficients is obtained by $\Bar{p}_i$ in case of the endpoint, and $\Delta qU \Bar{p}_i$ in case of the squared neutrino mass. Given the analysis range on the 10\,eV scale and scattering probabilities in the 10\,\% regime, $\epsilon_i$ is on the \SI{1}{eV} scale, such that the squared neutrino mass is very susceptible to $\Delta_{10}$.

\section{Supplements}
\label{ch:Supplements}

The following sections discuss further details, which are relevant for the previous discussions.

\subsection{Higher-order effects}
\label{sec:HigherOrders}

In this section several higher-order effects are discussed, that were neglected in the derivation of the main text.

\subsubsection*{Energy and angle dependency of the electron distributions}
\begin{figure}
    \centering
    \includegraphics[width=.49\textwidth]{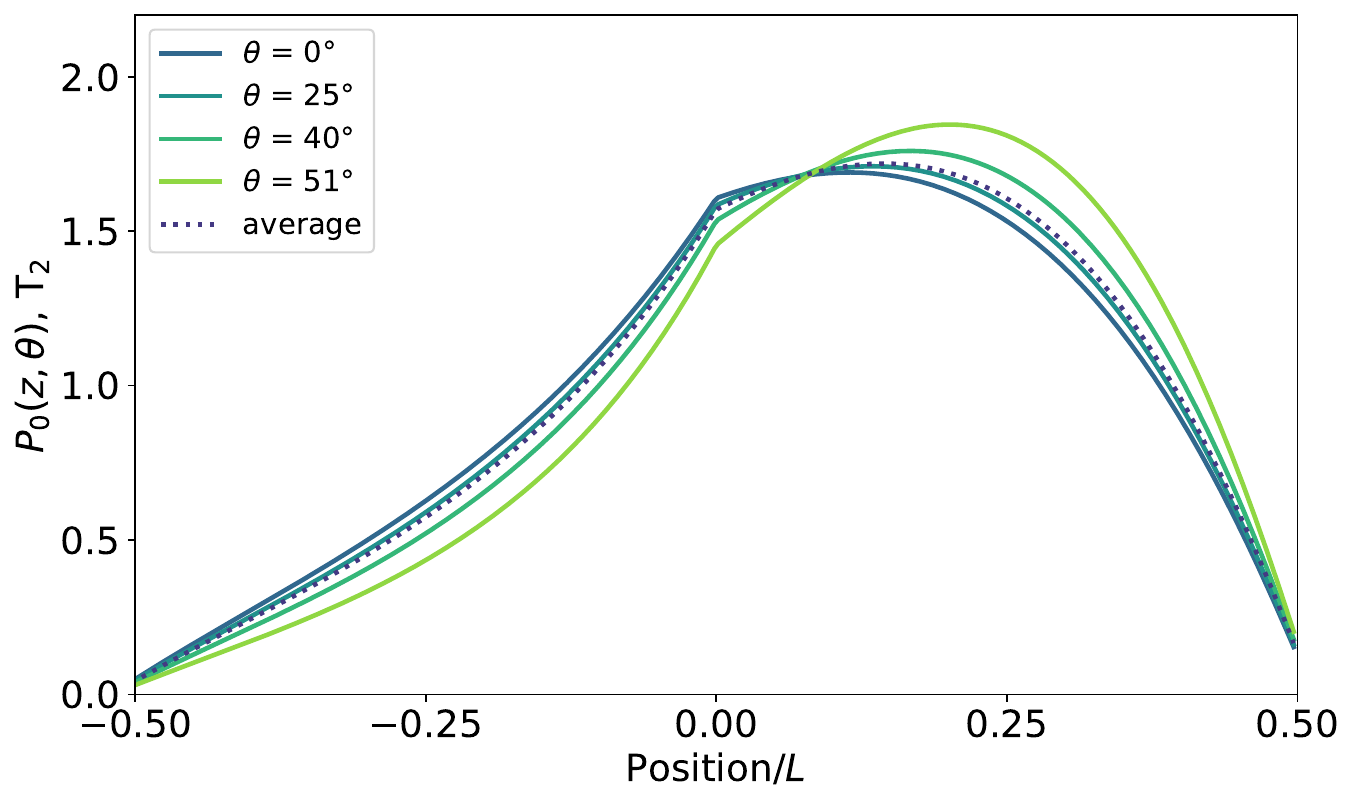}
    \caption{\textbf{Dependency of the unscattered electron distribution on the angle.}
    The electron distributions $P_0(\theta,z)$ are a function of $\cos{\theta}$. Accordingly, the change with $\theta$ increases, when $\theta$ is larger. Electrons with a large angle are more likely to stem from the front of the source, compared to electrons with a small angle.}
    \label{fig:ElectronDistributionsTheta}
\end{figure}
In the main text the energy and angle dependency of the electron distributions is neglected, and instead they are evaluated at a fixed energy $E_\mathrm{fix}$ and with scattering probabilities averaged over the angle.

The dependency on the energy is very small and also mitigated by the fact, that depending on the multiplicity of scattering, the initial energy of the electron can only deviate by a certain amount from $E_\mathrm{fix}$, otherwise the electron is scattered outside of the analysis range. This leads to a maximum spread of the initial energy of
\begin{equation}
    \Delta E = 40~\mathrm{eV}-i\times 13~\mathrm{eV}
\end{equation}
for an electron, that is scattered $i$-times. Here $40~\mathrm{eV}$ is a typical analysis range and $13~\mathrm{eV}$ the approximate minimum energy loss in one scattering event. Accordingly, unscattered electrons can have a spread in initial energy of $40~\mathrm{eV}$, while the (for the $40~\mathrm{eV}$ range) maximally allowed 3-times scattered electrons have a spread of only $1~\mathrm{eV}$.

The effect of the angle on the other hand, is more pronounced. Figure~\ref{fig:ElectronDistributionsTheta} shows the distribution of the unscattered electrons for $\tr$ depending on the angle. For large angles there is a visible change of the distributions, and thus of the sensitive region to the potential.

One can quantify the uncertainty resulting from the visible spread in the electron distribution using methods already developed in the main text. Using a Taylor expansion of the response function equation~\ref{eq:responsePotential} and calculating the maximum difference for the smallest and largest possible angle, the first order correction term is proportional to
\begin{align}
    \Delta_\theta[V]&\equiv\Braket{VP_0(\theta=\theta_\mathrm{max}}-\Braket{VP_0(\theta=0)}~, \\
    &\equiv\Braket{VP_{\Delta\theta}}~.
\end{align}
Analogous to section~\ref{sec:StatisticalMoments} and equation~\ref{eq:Inequality}, given a measured scale $\sigma_0[V]$ of the potential inhomogeneity, $\Delta_\theta[V]$ is bounded by
\begin{equation}
    |\Delta_\theta[V]|\leq\kappa\sigma_0[V]~.
\end{equation}
The potential, that produces the largest effect, is given by
\begin{equation}
    Q(z)=\frac{1}{\kappa}\frac{P_{\Delta\theta}}{P_0}(z)~,
    \label{eq:worstPotential}
\end{equation}
and the coefficient $\kappa\approx0.2$ is the standard deviation of $\frac{P_{\Delta\theta}}{P_0}$. Neglecting phase space factors in $\theta$, and assuming the unlikely case that the potential is finely tuned according to equation~\ref{eq:worstPotential}, this represents the maximum relative uncertainty of the first potential moments. While further studies are required to determine a fully conclusive value, it is plausible that this uncertainty is significantly overestimated.

Moreover, as demonstrated in this work, the primary sensitivity of the neutrino-mass measurement to the source potential originates from $\Delta_{10}[V]$. The effect discussed here constitutes an additional modification of the transmission edge, resulting in a neutrino-mass bias that is smaller by approximately an order of magnitude.

\subsubsection*{Inhomogeneity of the source magnetic field}

In principle, the electron distributions $P_i(z)$ must account for all $z$-dependent quantities that affect the rate of the electrons. One additional $z$-dependency worth mentioning arises from a slight decrease of the source magnetic field between the coils. Its impact on the electron rate is governed by two opposing effects: On the one hand, the magnetic field determines the volume of the flux tube; on the other hand, it sets the maximum pitch angle of the electrons, beyond which they are intentionally reflected by a strong magnetic field in front of the detector. As visible in figure~\ref{fig:ElectronDistributions}, the influence of the inhomogeneity of the source magnetic field on the electron distributions is very small, and can be neglected. Usually a constant, effective value is used, which is designed to compensate the neutrino-mass bias from the then neglected $z$-dependency~\cite{KATRINKNM1-5}.

\subsection{Parameterizing the $\Delta$-parameter space and coverage}
\label{sec:ParameterSpace}

Equation~\ref{eq:RhoVecEquation} is a quadratic equation with a positive and negative solution. For the first 3 scatterings the volume enclosed by the solution can be parametrized using parameters $r_i\in[-1,1]$ like
\begin{align}
    \hat{\rho}_1&=r_1~, \label{eq:rho1} \\
    \hat{\rho}_2&=\hat{\rho}_1\rho_{12}+r_2\sqrt{(1-\rho_{12}^2)(1-\hat{\rho}_1^2)}~, \label{eq:rho2}\\
    (1-\rho_{12}^2)\hat{\rho}_3&=\hat{\rho}_1(\rho_{23}\rho_{12}-\rho_{13})-\hat{\rho} _2(\rho_{13}\rho_{12}-\rho_{23})~, \\
    &+r_3\sqrt{|\korrM|\left[(1-\hat{\rho}_1^2)(1-\hat{\rho}_2^2)-(\rho_{12}-\hat{\rho}_1\hat{\rho}_2)^2\right]}~,
\end{align}
with determinant
\begin{equation}
    |\korrM|=(1-\rho_{13}^2)(1-\rho_{23}^2)-(\rho_{12}-\rho_{13}\rho_{23})^2~.
\end{equation}
The resulting $\vec{\rho}$ is translated to $\vec{\Delta}$ using equation~\ref{eq:ShapeOperator}, and those values are used to shift the convolved energy-loss function~\ref{eq:elossitimes}.

The usage of bounded parameters can produce complications regarding the treatment of uncertainties and the coverage. It has been shown in Monte Carlo studies that in order to obtain correct coverage, in addition to the necessary boundaries in the interval $[-1,1]$, a Gaussian pull term should be set on the $r_i$, centered at 0 and with a standard deviation of 1.

\subsection{Differences of $\tr$ and $\kr$ and measurement of $\Delta_{10}$}
\label{sec:KrT2Differences}

Due to the different gas profiles of $\tr$ and $\kr$, the moments of the starting potential that both spectra observe are different even for the same potential.

In the treatment of the $\Delta_{i0}$ the translation between the gas species is considered by calculating the covariances $\kappa_{ij}$ accordingly. In the KNM1\&2 measurement campaigns the $\tr$ and $\kr$ measurement were taken at different source conditions, possibly affecting the source potential. This is considered in \cite{KryptonMeasurementPaper} by drastically increasing the uncertainty of $\sigma_0$. It can then be interpreted as the value observed by the $\tr$ electrons, such that the limits on $|\Delta_{i0}|$ from equation~\ref{eq:Inequality} have to use the $\tr$ electron distribution $P_0(z)$ in the denominator of the $Q_i(z)$.

In all other campaigns, the $\tr$ and $\kr$ measurements are taken at the same source conditions. Since the neutrino-mass bias caused by the broadening is small, the scaling of $\sigma_0$ from $\kr$ to $\tr$ can be neglected. Also, as shown in~\cite{Mac21} the uncertainty in this scaling is larger, when the absolute value of $\hat{\rho}$ is smaller; in this case the neutrino mass bias is dominated by the component related to $\hat{\rho}$ anyway. Then, $\sigma_0$ from the $\kr$ measurement can directly be used in equation~\ref{eq:Inequality} to constrain the $\Delta_{i0}$, such that the distribution $P_0(z)$ of the $\kr$ electrons is used in the denominator of the $Q_i(z)$.

In all cases, the difference $P_i(z)-P_0(z)$ in the numerator of the $Q_i(z)$ is taken from the $\tr$ electron distributions, since the $\tr$ $\Delta_{i0}$ should be constrained. Rescaling the $\hat{\rho}$-ellipsoid to a $\Delta$-ellipsoid is always done using the $\kappa_i$ coefficients for $\tr$.

The measurement of $\Delta_{10}$ and $\sigma_0$ in $\kr$ allows to calculate $\hat{\rho}_1$ of krypton using equation~\ref{eq:ShapeOperator} and the value of $\kappa_{1,\mathrm{Kr}}$ for the $\tr$ column density used in the $\kr$ measurement. $\hat{\rho}_1$ is then used to constrain the respective value of tritium following from equation~\ref{eq:rho2} by
\begin{equation}
    \hat{\rho}_{1,\mathrm{T2}}=\hat{\rho}_{1,\mathrm{Kr}}\rho_{11,\mathrm{KrT2}}\pm\sqrt{\left(1-\rho_{11,\mathrm{KrT2}}^2\right)\left(1-\hat{\rho}_{1,\mathrm{Kr}}^2\right)}~,
    \label{eq:FromKrToT2}
\end{equation}
where the correlation $\rho_{11,\mathrm{KrT2}}$ between the $\kr$ and $\tr$ $P_1(z)-P_0(z)$ is used. This value of $\hat{\rho}_{1,\mathrm{T2}}$ is then the input of equation~\ref{eq:rho1}. In the case where the measurement fluctuates close to a boundary of the allowed parameter space, further coverage studies might be required. The $\kappa$ and $\rho$ values for the translation are shown in table~\ref{tab:CorrelationsKrT2}.
\begin{table}[h]
    \centering
    \caption{Standard deviation $\kappa_{1,\mathrm{Kr}}$ and correlation $\rho_{11,\mathrm{KrT2}}$ for the translation between the $\kr$ and $\tr$ measurement in the $75\,\%$ measurement.}
    \begin{tabular}{cccc}
    \toprule
         Measurement & Configuration & $\kappa_{1,\mathrm{Kr}}$ & $\rho_{11,\mathrm{KrT2}}$ \\
         campaign &  (CD, Temp) \\
    \midrule
        Since KNM3-NAP & \SI{75}{\percent}, \SI{80}{\kelvin} & 0.73 & 0.95 \\
    \bottomrule
    \end{tabular}
    \label{tab:CorrelationsKrT2}
\end{table}
Only the values for the $75\,\%$ measurement are shown, since here the same conditions are used for $\kr$ and $\tr$, such that the source potential can be assumed to be the same. If this is not given, then $\Delta_{10}$ cannot be translated between the two spectra without further assumptions, which is the case for all other measurement campaigns.

The limit on the $\tr$ $\Delta_{10}$ by equation~\ref{eq:Inequality} and the measured broadening $\sigma_0\approx 32~\mathrm{mV}$ is around $22\,\mathrm{mV}$~\cite{KryptonMeasurementPaper} for the last neutrino-mass campaigns. The projected uncertainty on the $\kr$ $\Delta_{10}$ is around $7\,\mathrm{mV}$, which would be a factor 3 smaller. Unfortunately, even though the $\tr$ and $\kr$ $\Delta_{10}$ are very highly correlated, the translation by equation~\ref{eq:FromKrToT2} adds a significant amount of uncertainty to the $\tr$ value. The largest increase happens, when the central value is 0, resulting in a symmetrical $\tr$ uncertainty of $13\,\mathrm{mV}$. If the central value of $\left|\Delta_{10}\right|$ is larger, for example $13\,\mathrm{mV}$, the resulting $\pm1\,\sigma$ interval of the tritium value is still $[-2\,\mathrm{mV},+22\,\mathrm{mV}]$, allowing to constrain approximately half of the parameter space. Only when the measured $\kr$ $\left|\Delta_{10}\right|$ is very large the size of the $\tr$ $\pm1\,\sigma$ interval approaches that of the $\kr$ measurement.

\subsection{Dependency of $z$-averaged quantities on the gas profile}

The covariances $\kappa_{ij}$ and correlations $\rho_i$ are important factors in quantifying the neutrino-mass bias caused by the source potential. They are obtained by calculating averages of the $z$-dependent electron distributions $P_i(z)$. For the case of $\tr$, these averages do not depend on the detailed shape of the density profile, but only on the total column density~\footnote{Thanks to F. Glück for pointing this out!}, making them robust against uncertainties in the modeling or in variations of source properties like the temperature.

This can be understood by noting that the scattering probabilities (c.f.\ equation~\ref{eq:ScatteringProbabilities}) depend only on the effective column density $\mathcal{N}_\mathrm{eff}(\theta,z)$, which itself is an integral over the density profile. Consequently, integrals over $z$ involving the density profile can be substituted by integrals over the effective column density.

However, in the case of $\kr$, this simplification does not apply. While the scattering probabilities are still governed by the $\tr$ density, the electron distributions $P_i(z)$ are directly proportional to the $\kr$ density.

\subsection{Effective treatment of the $\Delta$-parameter space}
 \label{sec:EffectiveDelta}

The first order bias of the squared neutrino mass is given by
\begin{equation}
    \Delta m^2=-\sum_{i=1} \epsilon_i \Delta_{i0}~.
    \label{eq:m2BiasLinear}
\end{equation}
If the susceptibilities $\epsilon_i$ are known, then an effective $\Delta_\rho$ can be defined like
\begin{equation}
    \Delta_\rho\equiv\frac{1}{\epsilon}\sum_{i=1} \epsilon_i \Delta_{i0}~,
    \label{eq:EffectiveDelta}
\end{equation}
such that the bias of the squared neutrino mass is given by
\begin{equation}
    \Delta m^2=-\epsilon \Delta_\rho~.
\end{equation}
The normalization $\epsilon$ determines how the effective $\Delta_\rho$ is used in the model. If all $\Delta_{i0}$ are shifted equally by $\Delta_\rho$, then it follows from equation~\ref{eq:m2BiasLinear} that $\epsilon$ is the sum of the $\epsilon_i$.

By construction, this approach gives the correct bias of the squared neutrino mass, but it has many disadvantages:
\begin{itemize}
    \item The $\epsilon_i$ have to be determined beforehand, which is usually done in Asimov studies neglecting correlations to other systematic effects and using a simplified analysis.
    \item Since the $\epsilon_i$ depend on the performed analysis, different values of $\Delta_\rho$ have to be used for different analyses. This will be especially noticeable when changing the analysis range, leading to a large change of $\Delta_\rho$.
    \item Even if the $\epsilon_i$ are correct for the given analysis, only the neutrino mass bias is compensated. The bias of the other observables and the residual structure caused by non-vanishing $\Delta_{i0}$ remains or can even be amplified.
\end{itemize}
Therefore, it is strongly recommended to treat the entire parameter space correctly, using the equations presented in section~\ref{sec:ParameterSpace}.

\section{Conclusions and outlook}

For KATRIN to reach its sensitivity goal, a careful assessment of systematic biases is required. This work investigated the bias caused by inhomogeneities in the source potential, demonstrating that the resulting effects are significantly larger than those accounted for in KATRIN’s predecessor experiments. These effects were discussed in leading order and the corresponding observables of the starting potential were derived. In particular, due to the existence of electron scattering, the leading order neutrino mass bias by the inhomogeneity of the potential is in contrast to the previous believe not caused by a Gaussian broadening $\sigma_0[V]$, but by mutual shifts $\Delta_{i0}[V]$ of the spectra of different scattering multiplicities. Since this effect produces residuals not at the endpoint, but starting from $\approx 13\,\mathrm{eV}$ deep into the spectrum, its possible neutrino-mass bias is many factors larger than expected simply by a Gaussian description. It was motivated in this work why the bias scales with approximately $\Delta m^2\propto 1\,\mathrm{eV}\cdot\Delta_{10}$. It was also shown that the parameter space of the $\Delta_{i0}$ is an ellipsoid constrained by the standard deviation $\sigma_0$ of the potential, and that the boundary of the ellipsoid is produced by source potentials that are asymmetric with regard to the injection point in the center of the source. To constrain the parameter space further, the $\kr$ measurement is used, where the effect of asymmetric potentials is directly visible as a shift between the 1x scattered and unscattered line. The experimental approach of the $\kr$ measurement is described in a parallel publication~\cite{KryptonMeasurementPaper}, where $\sigma_0^2=1.0(3)\times 10^{-3}\,\mathrm{eV}^2$ is determined, leading to a systematic neutrino-mass uncertainty on the order of $2\times 10^{-2}\,\mathrm{eV}^2$. In the future the already recorded measurement of $\Delta_{10}$ will be analyzed, and as shown in this work, it is foreseen that it will allow to reduce the systematic bias on the neutrino mass further by a factor of around 2 to 3, depending on the central value of the measured $\Delta_{10}$.

\begin{acknowledgements}
I would like to thank M. Böttcher, M. Kleesiek, M. Schlösser, M. Slez\'ak, A. Lokhov and K. Valerius for the helpful discussions and support. Furthermore I thank C. Fengler, J. \v{S}torek and F. Glück for the review of the manuscript.
\end{acknowledgements}

\bibliographystyle{unsrtnat}
\bibliography{references}

\end{document}